\documentclass[aps,prc,reprint,groupedaddress]{revtex4-1}

\newcommand{\LA}{$\Lambda(1405)$ }
\newcommand{\LAo}{$\Lambda(1405)$}
\newcommand{\KNN}{$\overline{K}NN$ }
\newcommand{\KNNo}{$\overline{K}NN$}
\newcommand{\pkl}{$pK^{+}\Lambda$ }

\newcommand{\pL}{$p\Lambda$ }

\newcommand{\MeV}{MeV/c$^{2}$ }
\newcommand{\MeVo}{MeV/c$^{2}$}
\newcommand{\ppk}{"$ppK^{-}$" }
\newcommand{\ppko}{"$ppK^{-}$"}

\usepackage{pst-all}                                         % ps tricks 
\usepackage{graphicx}                                     % Include figure files
\usepackage{amsmath}                                    % this is for flalign to aling formulars

\begin{document}

% Use the \preprint command to place your local institutional report
% number in the upper righthand corner of the title page in preprint mode.
% Multiple \preprint commands are allowed.
% Use the 'preprintnumbers' class option to override journal defaults
% to display numbers if necessary
%\preprint{}

%Title of paper
\title{
Has a "$ppK^{-}$" Signal in p+p Reactions been Observed Yet?\\
}

% repeat the \author .. \affiliation  etc. as needed
% \email, \thanks, \homepage, \altaffiliation all apply to the current
% author. Explanatory text should go in the []'s, actual e-mail
% address or url should go in the {}'s for \email and \homepage.
% Please use the appropriate macro foreach each type of information

% \affiliation command applies to all authors since the last
% \affiliation command. The \affiliation command should follow the
% other information
% \affiliation can be followed by \email, \homepage, \thanks as well.
\author{E.~Epple$^{1,2}$ and L.~Fabbietti$^{1,2}$}

\affiliation{
\mbox{eliane.epple@ph.tum.de, laura.fabbietti@ph.tum.de}
} 
%\email[]{Your e-mail address}
%\homepage[]{Your web page}
%\thanks{}
%Collaboration name if desired (requires use of superscriptaddress
%option in \documentclass). \noaffiliation is required (may also be
%used with the \author command).
%\collaboration can be followed by \email, \homepage, \thanks as well.
%\collaboration{}
%\noaffiliation

\date{\today}

\begin{abstract}
% insert abstract here
Our answer to the question posed in the title is: probably no.
In this work we show that it is rather unlikely that the structure $X(2265)$ reported by the DISTO 
collaboration corresponds to a kaonic nuclear bound state. 
The main argumentation is based on the repetition of the DISTO analysis on the HADES data sample, containing 
p+p reactions at 3.5 GeV. We further discuss many aspects in connection with the \pkl final state and the $\Lambda(1405)$-resonance. 
The results evidence possible problematics in the interpretation of the DISTO data.
\end{abstract}
%13.75.Jz    Kaon-baryon interactions
%13.75.Cs    Nucleon- Nucleon interactions
%25.40.-h	Nucleon-induced reactions 
%21.30.Fe	Forces in hadronic systems and effective interactions
%21.45.-v	Few-body systems
%%24.10.-i Nuclear reaction models and methods
%26.60.-c	Nuclear matter aspects of neutron stars
%21.65.-f	Nuclear matter
% insert suggested PACS numbers in braces on next line
%\pacs{13.75.Jz}\pacs{13.75.Cs}\pacs{25.40.-h}\pacs{21.30.Fe}\pacs{21.45.-v}\pacs{24.10.-i}\pacs{21.65.-f}
% insert suggested keywords - APS authors don't need to do this
%\keywords{}
%\maketitle must follow title, authors, abstract, \pacs, and \keywords
\maketitle

% body of paper here - Use proper section commands
% References should be done using the \cite, \ref, and \label commands
\section{The $\mathbf{pK^{+}\Lambda}$ final state in p+p reactions}
% Put \label in argument of \section for cross-referencing
%\section{\label{}}
Since the first measurement of open strangeness production via the reaction 
\begin{flalign}\label{eq:pkl}
 p+p  \rightarrow p  + K^{+}+ \Lambda  
\end{flalign}
has been reported \cite{PhysRev.108.1048}, many experiments have exclusively measured this final state. 
Three issues were mainly investigated thereby: 
the production of N$^{*}$ resonances and their subsequent decay into $K^{+}\Lambda$ 
\cite{Bierman:1966zz,Chinowsky:1968rn,Firebaugh:1968rq,Baksay:1976hb,Cleland:1984kb,AbdelSamad:2006qu,AbdElSamad:2010tz,AbdelBary:2010pc,Agakishiev:2014dha}, 
the \pL final state interaction \cite{Sibirtsev:2005mv,AbdelSamad:2006qu,Budzanowski:2010ib,Roder:2013gok}, and the cusp structure appearing at 2.13 GeV in the \pL invariant mass distribution \cite{Siebert:1994jy,Budzanowski:2010df,AbdEl-Samad:2013ida}. 
A fourth aspect was recently added to this list with the investigation of the kaonic nuclear bound state \KNN by the DISTO collaboration \cite{Yamazaki:2010mu,Maggiora:2009gh}.
In this analysis the following scenario has been investigated 
\begin{flalign}\label{eq:KNN}
\begin{pspicture}(0,0)
  \psline[ArrowInside=-]{->}(1.9,-0.10)(1.9,-0.45)(2.6,-0.45)
\end{pspicture}
 p+p  \rightarrow "ppK^{-}"& + K^{+}  \\ 
                                                 &\hspace{0.2cm} p+\Lambda .           \nonumber   
\end{flalign}
\par
Here, a kaonic nuclear bound state \KNNo, also called \ppko, is produced in p+p reactions together with a K$^{+}$, and its non-mesonic 
decay in a $p\Lambda$ pair has been considered. The aim of this work is to cross-check the claim 
that the observed structure (named $X(2265)$, M=2267\,\MeV and $\Gamma$=118\,MeV) in a so-called deviation spectrum of Refs.~\cite{Yamazaki:2010mu,Maggiora:2009gh} 
corresponds to the signature of an intermediate \KNN cluster to the final state~(\ref{eq:pkl}).
\par
In the following, we will explain what a deviation spectrum is, what difficulties arise from this method and why the absence of $X(2265)$ at 
lower beam energies is not linked to the absence of $\Lambda(1405)$ production - which is considered as a doorway for the formation 
of a \KNN~\cite{Yamazaki:2002uh,Yamazaki:2007cs}. 
We will further discuss whether a \KNN production strength of 17\% of the total \pkl production cross section at E$_{Kin}$=2.85\,GeV, as 
reported in Ref. \cite{Kienle:2011mi}, is a realistic scenario, given the 
upper limits for $X(2265)$ from Ref.~\cite{Kienle:2011mi} and this work. The discussion is completed by summarizing 
all \pL mass spectra published so far where no clear signature of $X(2265)$ or any other kind of \ppk signal is visible. 
%- - - - - - - - - - - - - - - - - - - - - - - - - - - - - - - - - - - - - - - - - - - - - - 
\section{The 2.5 GeV and 2.85 GeV discrepancy}
%- - - - - - - - - - - - - - - - - - - - - - - - - - - - - - - - - - - - - - - - - - - - - - 
After the publication of the DISTO results on the formation of a \ppk in p+p reactions at a beam kinetic energy of 2.85\,GeV 
\cite{Yamazaki:2010mu,Maggiora:2009gh}, the same authors have analyzed also a data set measured by DISTO at a beam kinetic energy of 2.5\,GeV \cite{Kienle:2011mi}. 
Despite the expectation, that as much as 33\% of the observed yield of the structure $X(2265)$ at 2.85\,GeV should be 
visible also at the lower beam energy of 2.5\,GeV, no signal appeared in the data.  
Therefore, an upper limit of 0.2$\pm$2.1\% of the \pkl production cross section was estimated \cite{Kienle:2011mi}.
\subsection{The depletion of the $\mathbf{\Lambda(1405)}$ yield}\label{LA-arg}
The missing signature of $X(2265)$ at 2.5\,GeV was explained by the lower abundance of the \LAo-resonance at this energy which is, according to 
Refs.~\cite{Yamazaki:2002uh,Yamazaki:2007cs}, a doorway for the formation of the \KNN in p+p reactions~\footnote{In this view the 
$\Lambda(1405)$, being partially a $p\overline{K}$-bound state, forms together with a proton a \KNN by final state interaction.}.  
The abundance of the \LAo-resonance was, however, only roughly estimated on the base of the missing mass distribution
to the proton and $K^+$ ($MM_{pK^{+}}$) at 2.85\,GeV and 2.5\,GeV. 
In this approach the high mass region of the $MM_{pK^{+}}$ spectra, which includes among others the contributions by the $\Lambda(1405)$ 
and $\Sigma(1385)^{0}$, was considered and the $\Lambda(1405)$ production at 2.5\,GeV was estimated to be maximally 10\% as for the 
data set at 2.85\,GeV \cite{Kienle:2011mi}. 
This estimation assumed that first, the statistic in the resonance region of the $MM_{pK^{+}}$ spectrum contains only the resonances $\Lambda(1405)$ and $\Sigma(1385)^{0}$, 
and second, that the $\Sigma(1385)^{0}$ to $\Lambda(1405)$ production ratio is the same for the two energies. 
The first assumption was disproven by the investigation of the $\Lambda(1405)$ resonance at a beam energy of 3.5\,GeV \cite{Agakishiev:2012xk}. 
Indeed in the latter work the individual contributions to the $MM_{pK^{+}}$ spectrum were determined and it was found that a substantial 
contribution stems from the production of the $\Lambda(1520)pK^{+}$, $\Sigma^{+}\pi^{-}pK^{+}$, and $\Delta^{++}
(1232)\Sigma^{-}K^{+}$ final states. 
While the $\Lambda(1520)$ production is below threshold at 2.5\,GeV ($E_{kin}$(Threshold) = 2.77\,GeV) and probably small at 2.85\,GeV, 
the other two final states will definitely contribute to the observed yield in the high mass region of the $MM_{pK^{+}}$, spectra measured by 
DISTO.  
The second assumption about a constant $\Sigma(1385)^{0}$ to $\Lambda(1405)$ production ratio is also questionable 
as the analysis in Ref.~\cite{Agakishiev:2012ja} showed that at 3.5\,GeV this ratio is reduced by about 15\% in comparison to the ratio measured 
by the ANKE collaboration at $E_{kin}$=2.83\,GeV \cite{Zychor:2007gf}. 
Both values have, however, large uncertainties so that it is difficult to extrapolate to lower energies. 
\par
\begin{figure}[t]
\begin{center}
\includegraphics[width=0.5\textwidth]{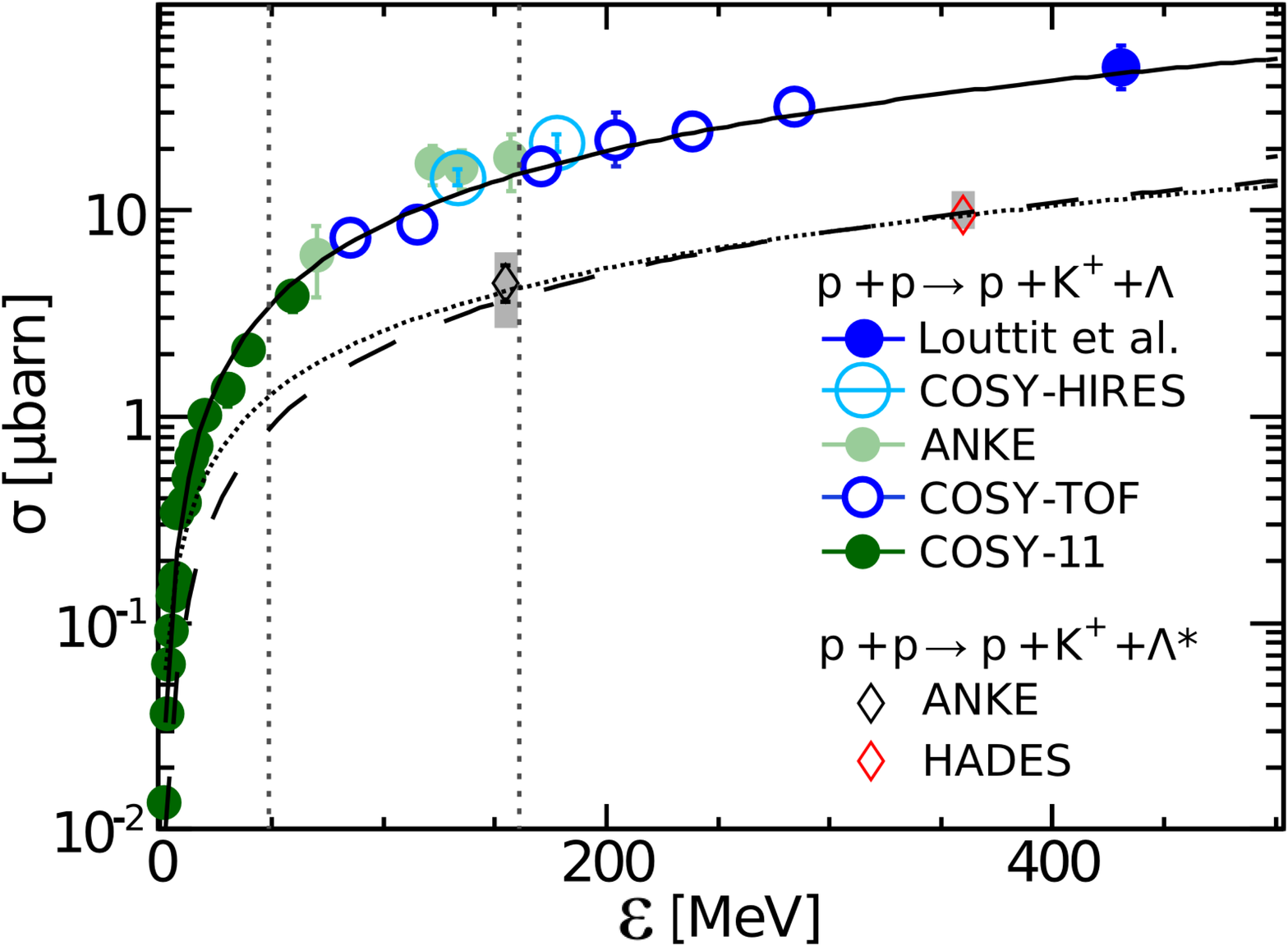}
\caption[]{(Color online) Compilation of measured $pK^{+}\Lambda$ (circles) and $pK^{+}\Lambda^*$ (diamonds) cross sections. 
The parametrisation that describes the $pK^{+}\Lambda$ data is taken from Refs.~\cite{AbdelBary:2010pc,Faldt:1996rh}. 
The long dashed curve shows this parametrisation scaled down to fit to the two $pK^{+}\Lambda^*$ data points and
 the dotted curve is a free fit of the F\"aldt-Wilkin parametrisation to the two points~\cite{Agakishiev:2012xk,Zychor:2007gf}. 
 The vertical lines show the $pK^{+}\Lambda^*$ excess energy for the two DISTO data sets.}
\label{Fig:La1405_CS}
\end{center}
\end{figure}

We suggest an alternative ansatz to compare the $\Lambda(1405)$ production cross section at the two DISTO energies. 
In Ref.~\cite{Agakishiev:2012xk} the energy dependence of the production cross section of the $pK^{+}\Lambda(1405)$ final state was 
determined on the base of the values measured in p+p collisions \cite{Agakishiev:2012xk,Zychor:2007gf}. Figure~\ref{Fig:La1405_CS} shows 
a compilation of measured cross sections from the $pK^{+}\Lambda$ and 
$pK^{+}\Lambda^*$ final states versus the excess energy. % for the particle production. 
The two vertical dashed lines mark the excess energy for the $\Lambda(1405)$ production for the two data sets, measured by DISTO 
(48.8\,MeV and 161.2\,MeV). 
The $pK^{+}\Lambda$ data are well described by a F\"aldt and Wilkin parametrisation as done in Refs.~\cite{AbdelBary:2010pc,Faldt:1996rh}. 
By assuming a similar behavior of the two channels close to threshold, we have scaled the \pkl curve down so that it fits the data points of 
the $pK^{+}\Lambda^*$ final state (long-dashed curve). 
We did also perform a free fit of the mentioned parametrisation from Refs.~\cite{AbdelBary:2010pc,Faldt:1996rh} to the two data 
points which also describes them well (dotted curve). 
With help of the two curves the ratio of the $\Lambda^*$ production cross section between the two DISTO energies was determined to be 
$\sigma_{pK^{+}\Lambda(1405)}(2.5\,GeV)$/$\sigma_{pK^{+}\Lambda(1405)}(2.85\,GeV)$=0.23, for the scaled curve and 0.3 for the curve 
based on the free fit to the data. The ratios show that the cross section of $\Lambda(1405)$ production at the 2.5\,GeV data 
set is in any case a sizable fraction of that at 2.85\,GeV. 
Following the assumption that the \KNN production in p+p collision should proceed through the intermediate 
formation of a $\Lambda(1405)$, at least 23\% of the observed $X(2265)$ yield at 2.85\,GeV should 
be expected at the lower energy. 
\par
In fact, the fraction of events affected by the $\Lambda(1405)p$ final state interaction should be even higher at lower 
energies due to phase space considerations \cite{Sibirtsev:2005mv}. 
Provided that the hypothesis of the \LA being a doorway for the creation of \ppk is valid, that would result in an increased number of \KNN per $\Lambda(1405)$.
\par
%And, thus, there are enough reasons to believe that there are \textbf{enough} grundvorraussetzungen for the formation of the bound state.
We, thus, argue that the reasoning in Ref.~\cite{Kienle:2011mi} regarding the absence of a $X(2265)$ signal at the 2.5\,GeV DISTO data 
is not convincing as the $\Lambda(1405)$ yield at lower energies is larger than estimated by the authors. 
%
%- - - - - - - - - - - - - - - - - - - - - - - - - - - - - - - - - - - - - - - - - - - - - - 
\subsection{The problematic of deviation spectra}
%- - - - - - - - - - - - - - - - - - - - - - - - - - - - - - - - - - - - - - - - - - - - - - 
%
%
To provide a further cross check of the reported results by the DISTO collaboration, we have repeated 
the analysis of Refs.~\cite{Yamazaki:2010mu,Maggiora:2009gh,Kienle:2011mi} and produced so-called deviation spectra. 
The original idea behind this approach was that any measured event distribution in a given observable which deviates 
from a purely phase space driven production process hints to
the presence of a new signal. The deviation spectrum is obtained by dividing the experimental event distribution of an observable by the same 
simulated distribution obtained by employing phase space simulations of the final state.
\par
\begin{figure}[t]
\begin{center}
\includegraphics[width=0.468\textwidth]{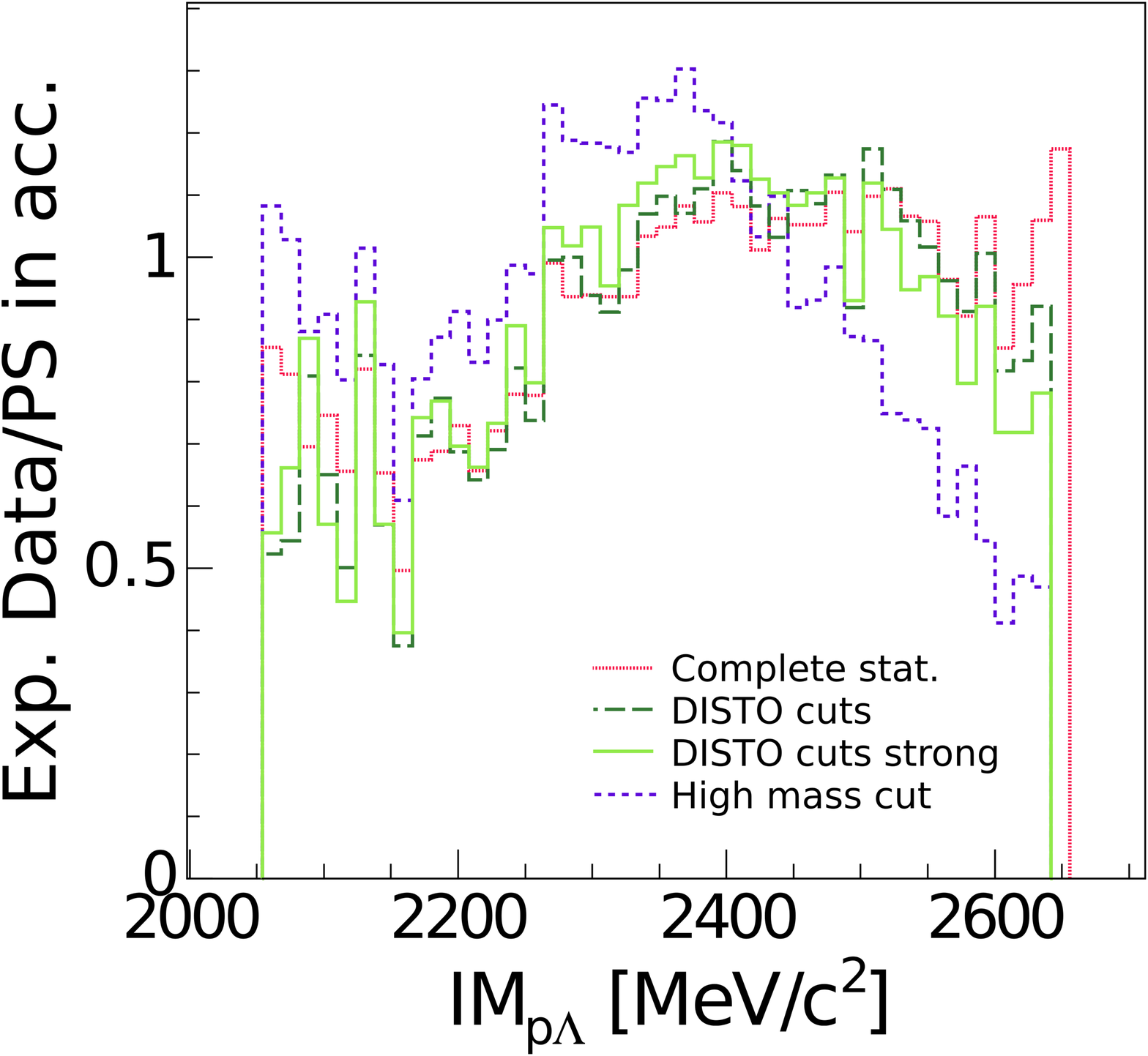}
\caption[]{(Color online) Different deviation spectra, obtained by dividing the reconstructed \pkl statistic by phase space simulations in the HADES acceptance.}
\label{Fig:DevHADES}
\end{center}
\end{figure}
\begin{figure}[t]
\begin{center}
\includegraphics[width=0.45\textwidth]{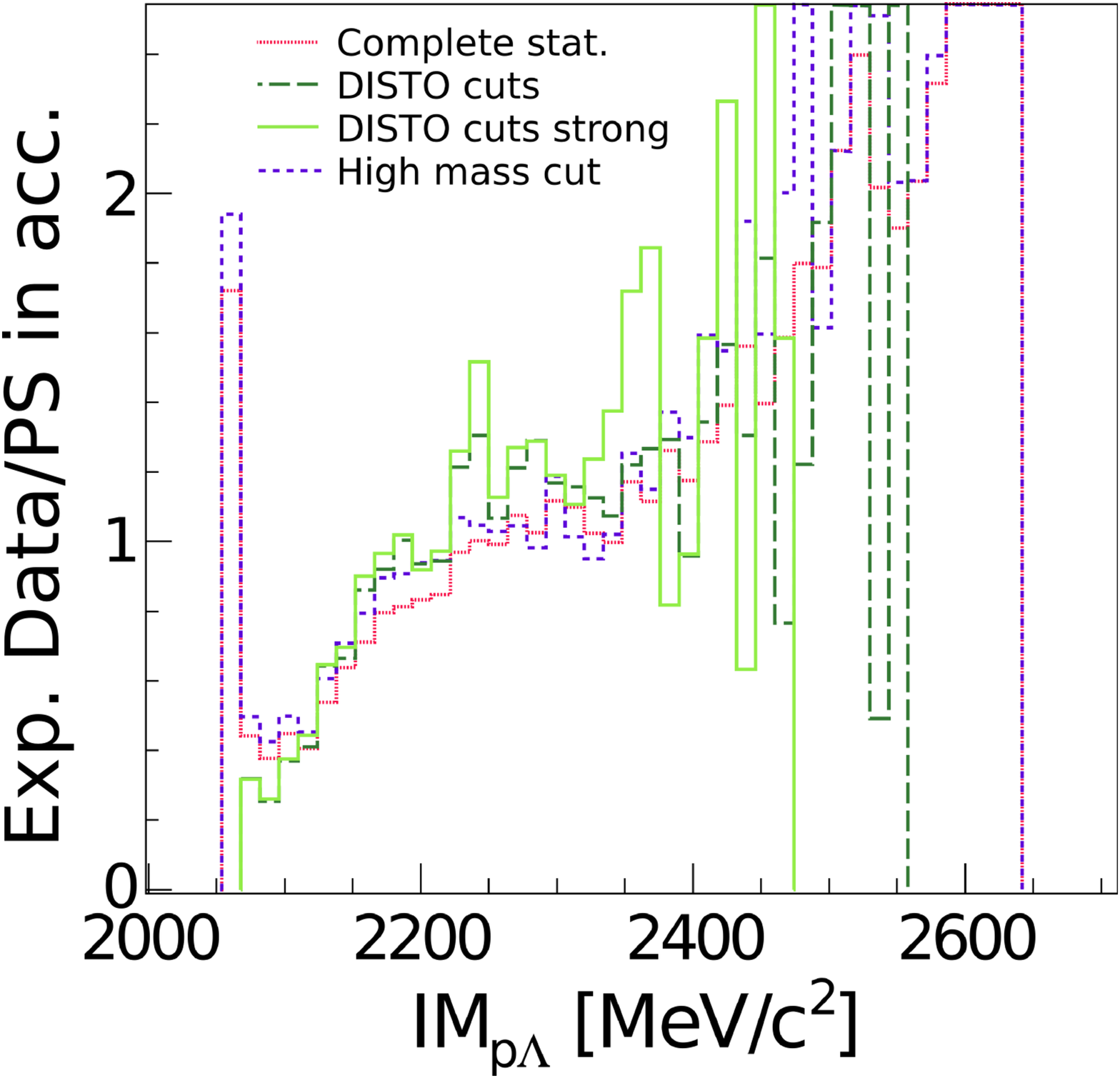}
\caption[]{(Color online) Different deviation spectra, obtained by dividing the reconstructed \pkl statistic by phase space simulations in the WALL acceptance.}
\label{Fig:DevWALL}
\end{center}
\end{figure}

The data for this analysis \cite{Epple:2012cq,Fabbietti:2013npl,Epple:2014iza,Agakishiev:2014dha} stem from the p(E$_{Kin}$=3.5\,GeV)+p 
reaction measured by the \textbf{h}igh-\textbf{a}cceptance \textbf{d}i-\textbf{e}lectron \textbf{s}pectrometer (\textbf{HADES}) at the SIS18 
synchrotron (GSI Helmholtzzentrum in Darmstadt, Germany). For details about the spectrometer and the experiment 
see Refs.~\cite{Agakishiev:2009am,Agakishiev:2014dha}. 
The in the following discussed deviation spectra were obtained by dividing the measured event distribution of the invariant mass distribution of p-$\Lambda$-pairs ($IM_{p\Lambda}$) by the 
according simulated spectrum. 
The simulations were performed under the assumption that the three particles in the final state (\ref{eq:pkl}) are produced via phase space 
only. 
The division is performed with measured spectra which means that the data are shown inside of the HADES acceptance 
and are filtered by the event selection procedure. 
As the \pkl statistic was obtained within two different regions of the spectrometer acceptance (HADES and WALL, see Ref.~\cite{Agakishiev:2014dha} 
for details) we present two different deviation spectra in Figs. \ref{Fig:DevHADES} and \ref{Fig:DevWALL}, respectively. 
\par
The two figures show several deviation spectra obtained under different data selections. While the red 
histograms show the deviation spectra for the full statistics as analysed in \cite{Agakishiev:2014dha}, the long-dashed histogram represents 
the result after applying the very same cuts as done in the analysis by DISTO: 
$\vert cos\theta_{p}\vert < $ 0.6 and -0.2$<cos\theta_{K^{+}}<$0.4 \cite{Yamazaki:2010mu,Maggiora:2009gh}. 
To point out the impact of such subsequent cuts, we have further restricted the proton 
angle from 0.6 to $\vert cos\theta_{p}\vert < $0.4, while leaving the cut for the kaons unchanged; this is illustrated by the green histograms. 
An remarkable result (violet dashed in Fig. \ref{Fig:DevHADES}) is obtained if one only selects events where $M_{K^{+}\Lambda}>$1810\,
\MeVo.
\par
Before interpreting these spectra it has to be noted that: 
first, it was shown already that phase space production is not a suited model to describe the \pkl final state at a beam energy of 3.5\,GeV 
\cite{Epple:2012cq,Fabbietti:2013npl}, so we do a priori not expect to retrieve a flat deviation spectrum from this method and secondly, 
most importantly, if the data are not described well by the simulation any further applied cut might have different influence on the measured 
and the simulated spectra. 
This is the reason why the deviation spectra drastically change under different cut conditions.
One can see, in particular for the HADES case, that the applied cuts start to deplete the deviation spectra from 2500\,\MeV on. As a 
consequence, one obtains a 
deviation spectrum with a rising slope up to 2400\,\MeV and a fall off for larger masses. 
This spectrum could be seen as broad peak structure, but it reflects only the effect of the kinematical cuts. 
Such an effect is as well observed for the WALL data-set in Figure \ref{Fig:DevWALL}, where the green dashed histogram shows a small structure at 2250\,\MeVo.
\par
\begin{figure}[b]
\begin{center}
\includegraphics[width=0.468\textwidth]{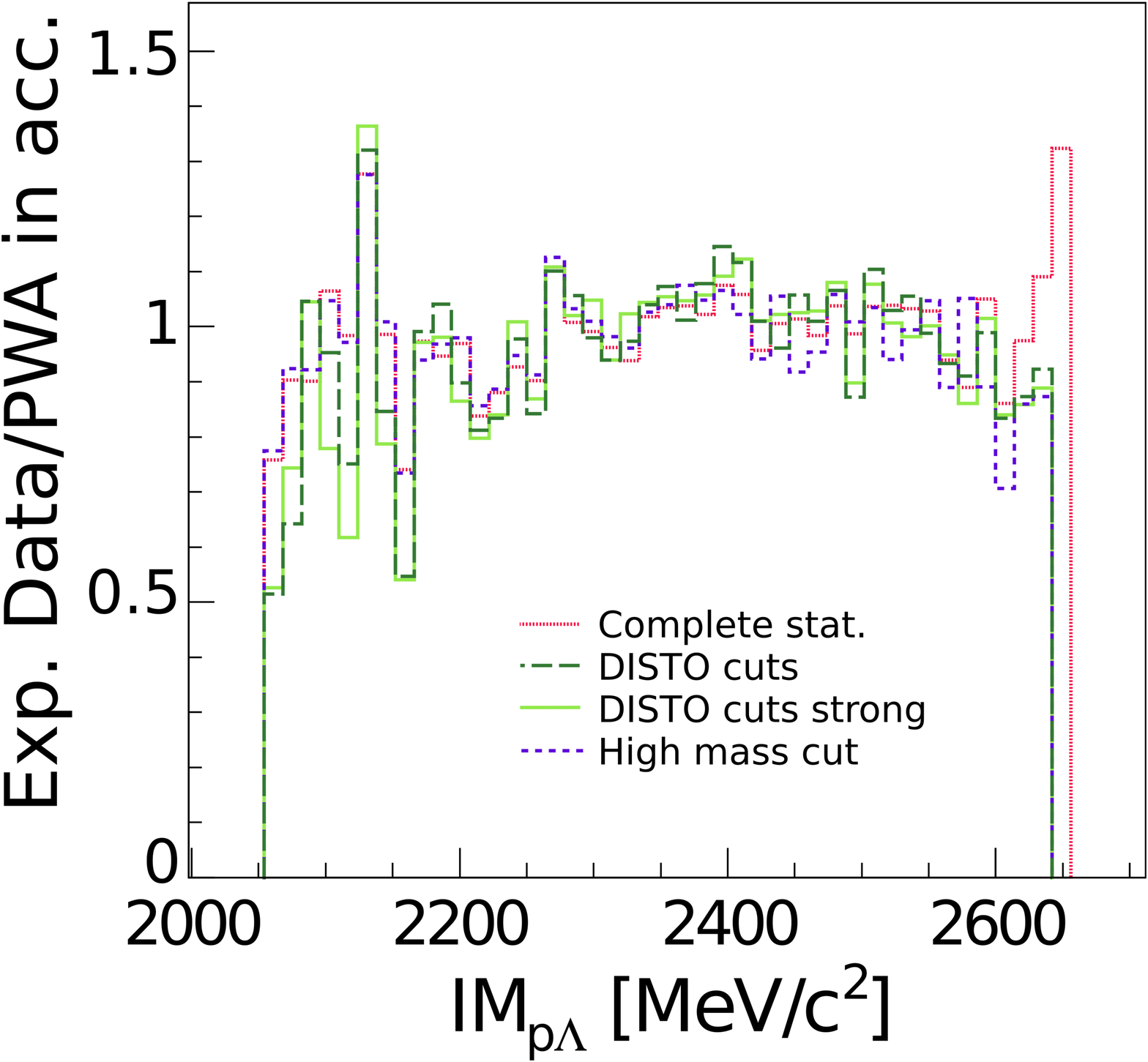}
\caption[]{(Color online) Different deviation spectra, obtained by dividing the reconstructed \pkl statistic by a partial wave analysis model \cite{Agakishiev:2014dha} in the HADES acceptance.}
\label{Fig:DevPWA}
\end{center}
\end{figure}

In order to clarify these observations we want to present deviation spectra that we have obtained by dividing the measured spectra by a partial wave analysis model 
\cite{Anisovich:2006bc,Anisovich:2007zz}.
This model is based on a partial wave analysis of the measured \pkl data \cite{Agakishiev:2014dha} and was compared to the measured events in many observables 
to gain confidence in the solidness of its data description. 
The deviation spectra in Figure \ref{Fig:DevPWA} are shown under the very same cut conditions as for the spectra with the phase space 
\begin{figure*}[t!]
\begin{center}
\includegraphics[width=0.8\textwidth]{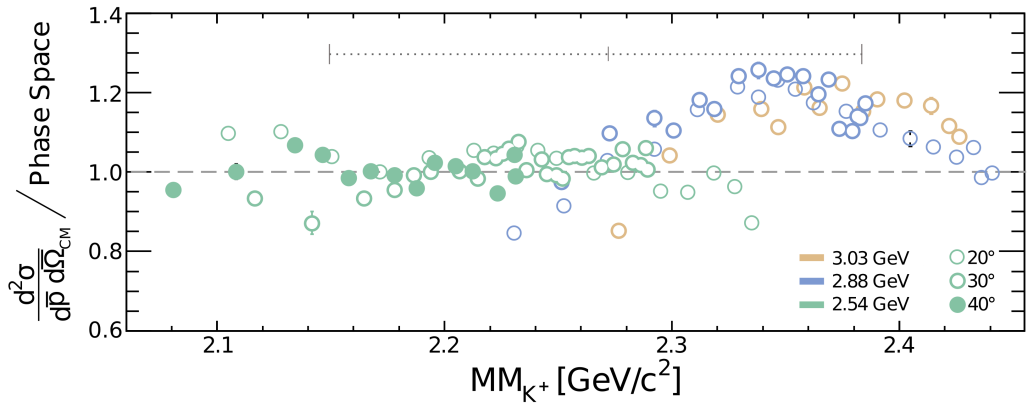}
\caption[]{(Color online) Spectra of the missing mass to the K$^{+}$, measured inclusively in p+p reactions at three different beam kinetic energies \cite{Hogan:1968zz}. 
The experimental spectra were divided by phase-space simulations. The gray horizontal bar indicates the range of the signal of $X(2265)$.}
\label{Fig:DevSpectra_Hogan}
\end{center}
\end{figure*}
simulations. In contrast to the Figs. \ref{Fig:DevHADES} and \ref{Fig:DevWALL}, the deviation spectra 
are in this case rather flat around one and the shape does not change when additional cuts are applied. 
This is entirely due to the correct description of the data by the PWA \cite{Agakishiev:2014dha} so that the applied cuts, therefore, 
act symmetrically on the data and the model.
\par
Summarizing, we have shown that deviation spectra strongly depend on the model to which the 
data are compared to. If the simulation model is not fully adequate, the applied cuts may distort the deviation 
spectrum drastically and no reliable conclusions can be drawn. We thus consider this method as suboptimal
to look for peak structures. 
From our perspective it is, thus, not astonishing that a certain structure at 2.85\,GeV can not be retrieved at the 2.5\,GeV DISTO data set. 
%- - - - - - - - - - - - - - - - - - - - - - - - - - - - - - - - - - - - - - - - - - - - - - 
\section{Deviation Spectra Revisited}
%- - - - - - - - - - - - - - - - - - - - - - - - - - - - - - - - - - - - - - - - - - - - - - 
%
We want to extend our discussion on deviation spectra to other energies. Indeed, the idea of comparing phase space simulation to 
experimental data for the Reaction (\ref{eq:pkl}) dates back to Ref.~\cite{Hogan:1968zz}.
There inclusive spectra of the missing mass to the kaon were divided by phase space distributions at different kaon angles and beam kinetic 
energies. 
The result of \cite{Hogan:1968zz} is shown in Figure~\ref{Fig:DevSpectra_Hogan}. The deviation spectrum differs from unity in the high 
missing mass region (the missing mass to the kaon (MM$_{K^{+}}$) corresponds to the mass of the residue X with which it is produced, 
e.g. X= \pL or $\Sigma N \pi$). 
The horizontal gray line indicates the signal range of $X(2265)$. 
No deviation at $MM_{K^+}$=2267\,\MeV is visible in these data sets. 
Indeed, the authors of Ref.~\cite{Hogan:1968zz} have investigated the deviations in 
Figure~\ref{Fig:DevSpectra_Hogan} under the assumption of a di-baryon being produced together with the kaon. They have, however, also considered 
the fact that the presence of N$^{*}$ resonances in the data might cause the observed deviations from phase space, as was already 
suggested by an earlier work \cite{Bierman:1966zz}. 
%
%
%
%
%- - - - - - - - - - - - - - - - - - - - - - - - - - - - - - - - - - - - - - - - - - - - - - 
\section{Any Signs of a Large Visible Signal?}
%- - - - - - - - - - - - - - - - - - - - - - - - - - - - - - - - - - - - - - - - - - - - - - 
%
Since no $X(2265)$ signal at the 2.5\,GeV DISTO data has been found, an upper limit for its production strength of 0.2$\pm$2.1\% of the
total \pkl production cross section has been estimated \cite{Kienle:2011mi}. 
An independent analysis of p+p data measured by HADES has also set an upper limit on the production of a \KNN in p+p 
reactions at $E_{kin}$=3.5\,GeV \cite{Agakishiev:2014dha}. 
The major difference between the DISTO and the HADES analysis is that in the former case visible bumps in the \pL invariant mass 
spectrum were associated to a signal, while the latter analysis is done with help of a partial wave analysis (PWA) \cite{Anisovich:2006bc,Anisovich:2007zz}. In the PWA the amplitude 
strength of a wave associated to the production of a kaonic nuclear bound state was determined and coherently added to all other contributing waves. 
These two different approaches prevent, however, a direct comparison of the signal strength and the upper limits. 
In the best case, a partial wave analysis would be performed on the three data sets simultaneously. For the time being, we have performed a 
simple incoherent analysis of the data to extract the upper limit of a \textit{visible signal strength}, in order to be able to compare it consistently 
to the DISTO results. 
Such an extracted limit does not necessarily correspond to the real signal strength which is maximally compatible with the data, 
as interferences with other signals are neglected in this approach. 
%
%- - - - - - - - - - - - - - - - - - - - - - - - - - - - - - - - - - - - - - - - - - - - - - 
\subsection{An incoherent upper limit for the $\mathbf{\overline{K}NN}$ production at 3.5\,GeV}
\begin{figure}[t]
\begin{center}
\includegraphics[width=0.41\textwidth]{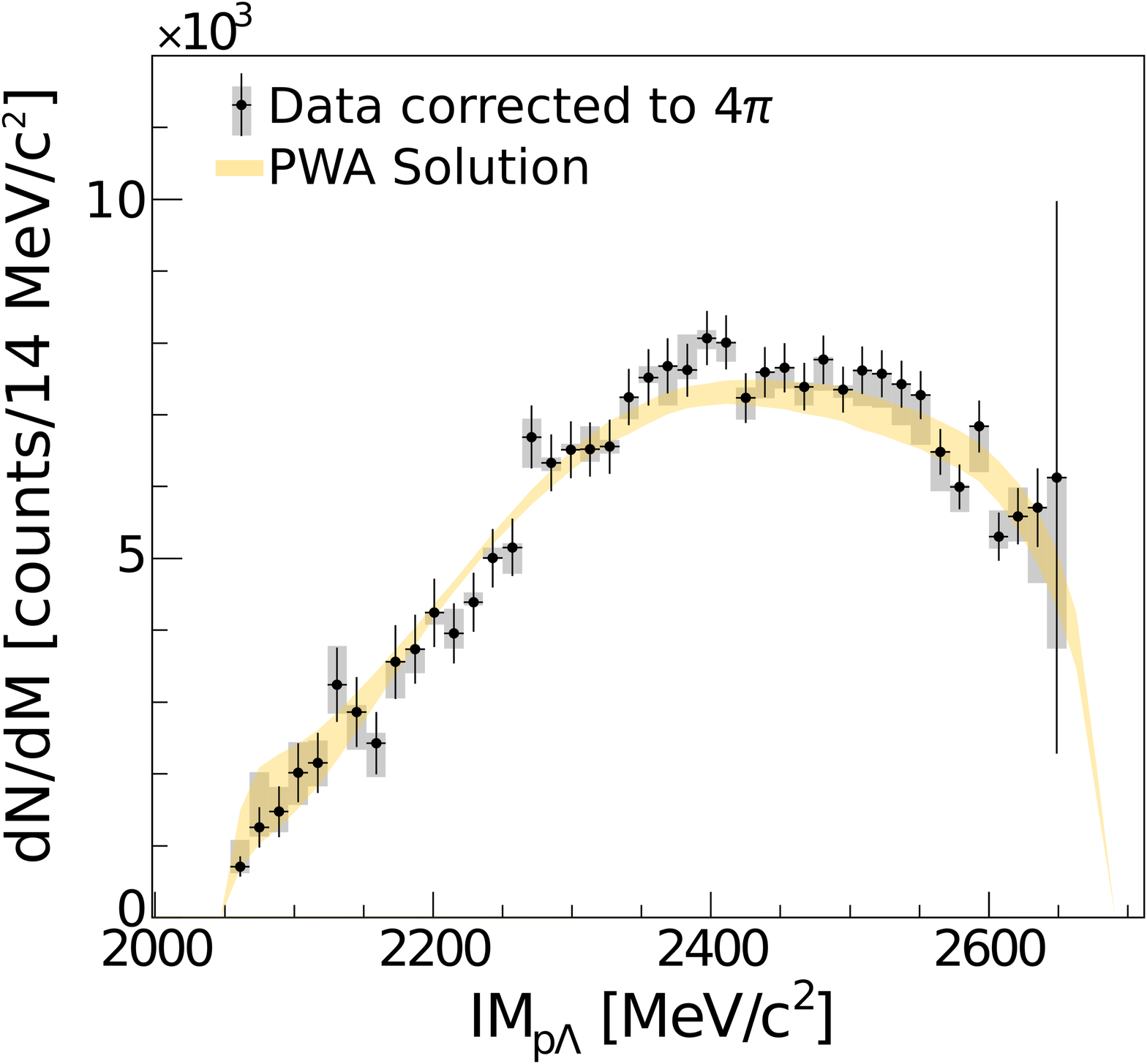}
\caption[]{(Color online) Acceptance  and efficiency corrected \pL invariant mass distribution in 4$\pi$ \cite{Epple:2014iza}. The yellow band shows the 4$\pi$ solutions from the PWA analysis without the inclusion of a kaonic nuclear cluster \cite{Agakishiev:2014dha}.}
\label{Fig:4piLP}
\end{center}
\end{figure}

\begin{figure}[t]
\begin{center}
\includegraphics[width=0.44\textwidth]{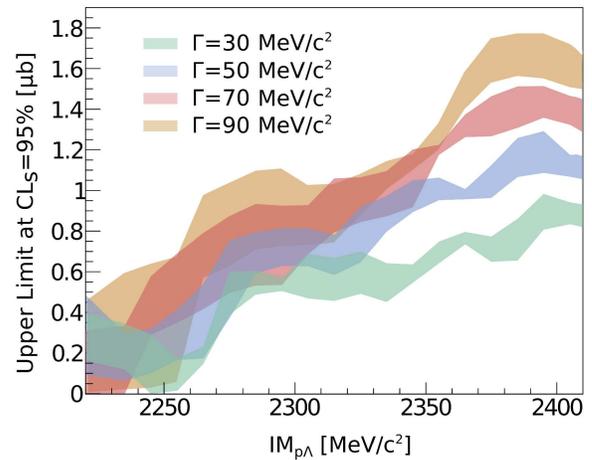}
\caption[]{(Color online) The incoherent upper limit as a function of the \KNN mass for four different widths of the bound state. The spread of the limit (widths of the bands) comes from the different results which are obtained from the four best solutions of the PWA \cite{Agakishiev:2014dha}.}
\label{Fig:UppLimit}
\end{center}
\end{figure}

%- - - - - - - - - - - - - - - - - - - - - - - - - - - - - - - - - - - - - - - - - - - - - - 
To carry out this analysis, we have used the acceptance corrected \pL invariant mass spectrum. 
The PWA solution from Ref.~\cite{Agakishiev:2014dha} delivers a model that 
describes the experimental spectrum well without the inclusion of a \KNN cluster 
signal. Although the PWA analysis was performed within the detector acceptance, 
an extrapolation of the solution to the full solid angle is possible.
This way, with help of the 4$\pi$-model also the measured data can be corrected to the full solid angle \cite{Epple:2014iza}. 
Figure~\ref{Fig:4piLP} shows the reconstructed experimental data in 4$\pi$ overlayed with the PWA solution. 
To extract an upper limit we have added a Breit-Wigner signal with varying mass, width, and amplitude to the PWA solution (signal 
hypothesis)
and compared the new spectra to the data. We have considered the model compatible with the data, if the 
confidence level of the signal strength ($CL_{s}$ \cite{Junk:1999kv,Read:2000ru,Read:2002hq}) was smaller than 95\%. 
Since there are small uncertainties from the acceptance correction (gray error bars in Fig.~\ref{Fig:4piLP}), 
the upper limit was determined four times. Each time the signal hypothesis was compared to the experimental data which were corrected 
with one of the four best PWA models of Ref.~\cite{Agakishiev:2014dha}. 
The resulting upper limit, as a function of the mass of the bound state, is shown in Figure \ref{Fig:UppLimit} for four different signal widths 
(30, 50, 70 and 90\,\MeVo). The width of the curves is due to the different upper limits obtained with the four different PWA models. 
The upper limit of about 0.7\,$\mu$b (in the relevant mass range) is below the coherent upper limit of the PWA of Ref.~\cite{Agakishiev:2014dha} which is consequential, as the interferences included in the PWA may hide the signal which makes it possible to include more signal strength while the spectra stay smooth.  
\par
We have also determined an upper limit specifically for the $X(2265)$ properties (M= 2267\,\MeVo, $\Gamma$= 118\,\MeVo) which is 0.3-1\,$\mu$b depending on the PWA solution 
to which the signal is added. 
\par
\begin{figure}[b]
\begin{center}
\includegraphics[width=0.45\textwidth]{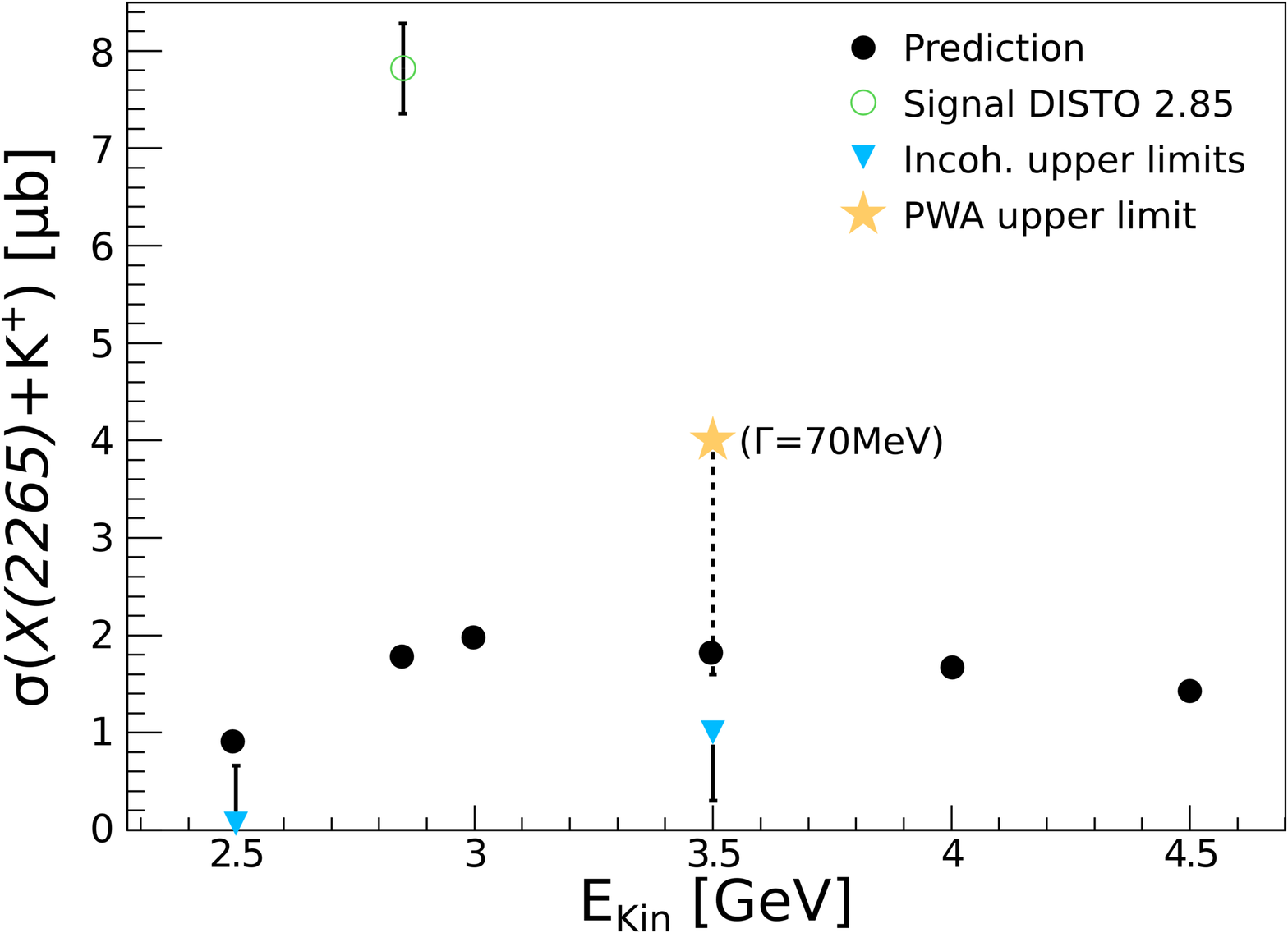}
\caption[]{(Color online) Predicted cross sections of a kaonic nuclear cluster production from Ref.~\cite{Suzuki:2009zze} (filled circles). 
The signal strength of the DISTO observation (open circle) and the upper limits from DISTO and HADES (triangle) given for the properties of $X(2265)$. 
The coherent upper limit from a PWA analysis \cite{Agakishiev:2014dha} (star) is given at the $X(2265)$ mass but for a width of 70\,MeV.}
\label{Fig:CrossS_KNN}
\end{center}
\end{figure}
Figure~\ref{Fig:CrossS_KNN} summarizes the observed yield at the DISTO energy of 2.85\,GeV, the coherent upper limit from Ref.~\cite{Agakishiev:2014dha} at 3.5\,GeV, 
the upper limit from the 2.5\,GeV DISTO measurement, the here extracted upper limit from the 3.5\,GeV HADES data set and the calculated prediction of a production cross 
section for the \KNN cluster of Ref.~\cite{Suzuki:2009zze}. 
While the upper limits for a visible signal strength at 2.5 and 3.5\,GeV lie apparently below the predictions of Ref.~\cite{Suzuki:2009zze} 
the DISTO value does exceed it by a factor of 4. 
No clear explanation for the excess with respect to the prediction at this specific energies was made available so far. 
This means in conclusion that either the cross sections for the production of $X(2265)$ has a very strong energy 
dependence or the observed structure $X(2265)$ is a non-physical signal. 
%
%- - - - - - - - - - - - - - - - - - - - - - - - - - - - - - - - - - - - - - - - - - - - - - 
\subsection{Qualitative Observations at Different p+p Beam Kinetic Energies}
\begin{figure*}[h]
\begin{center}
\includegraphics[width=0.8\textwidth]{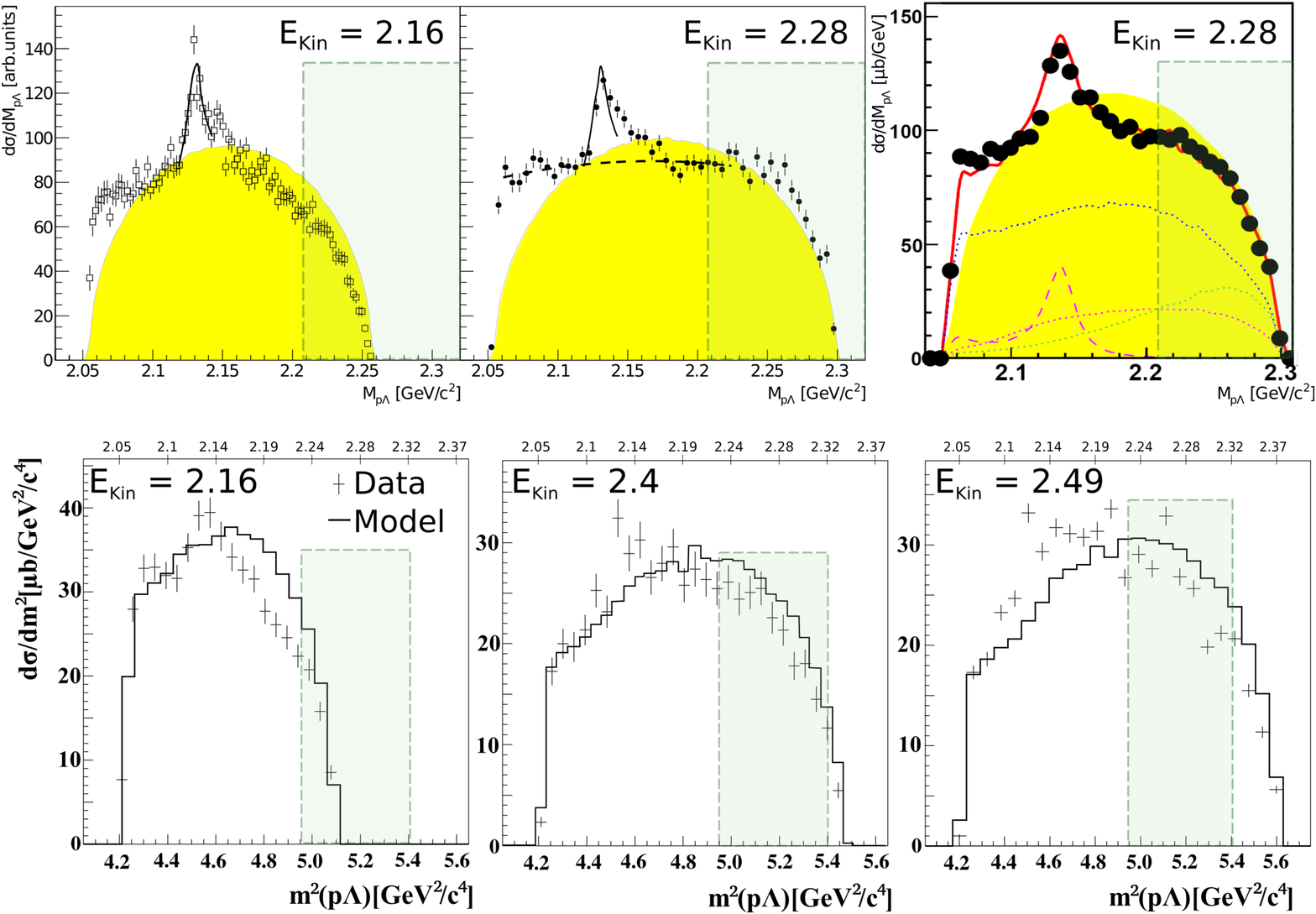}
\caption[]{The upper row shows measurements from the COSY-TOF Collaboration at two different beam kinetic energies \cite{AbdEl-Samad:2013ida}. 
The data are compared to phase space simulations indicated in yellow. The lower row shows also data from the COSY-TOF Collaboration at slightly higher beam kinetic energies \cite{AbdElSamad:2010tz}. The measured data are compared to a model that includes N*-resonances and final-state interaction. The green boxes mark the range of the $X(2265)$ signal (M$\pm\Gamma/2$).}
\label{Fig:PL_1}
\end{center}
\end{figure*}

\begin{figure*}[h]
\begin{center}
\includegraphics[width=0.9\textwidth]{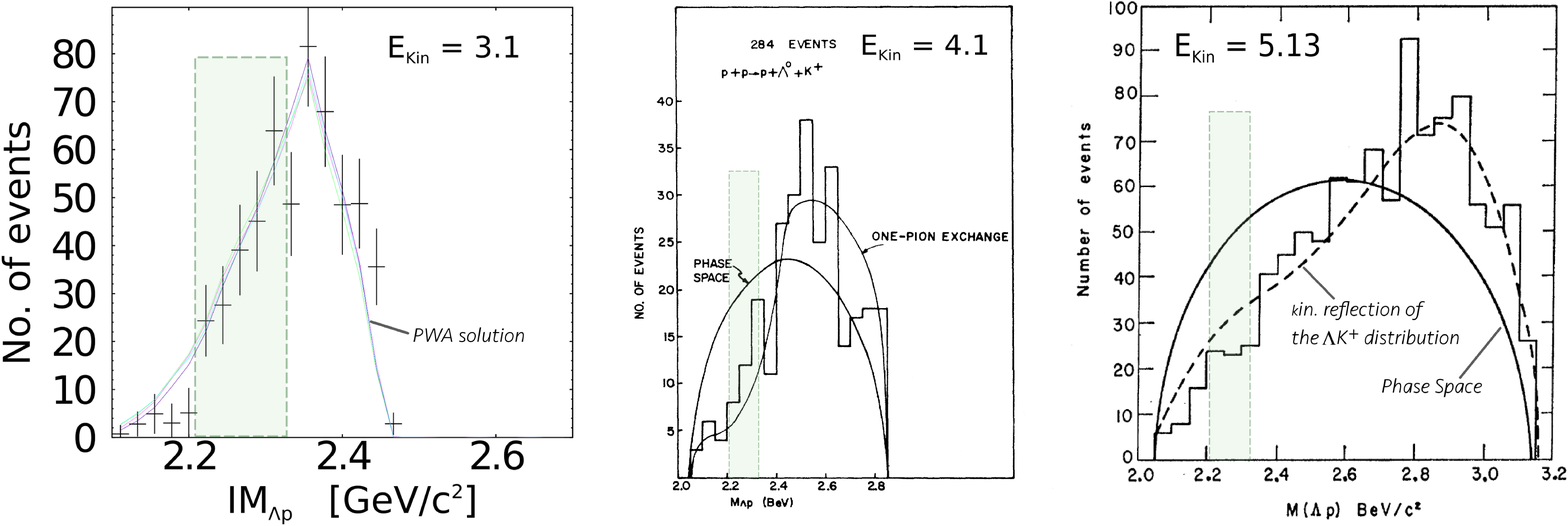}
\caption[]{The tree figures show data at a higher $E_{Kin}$ than used for the DISTO data sets. 
The left figure shows data measured by FOPI \cite{FOPI_Muenzer}. They are compared to solutions from a partial wave analysis. 
The middle figure shows data from the Brookhaven bubble chamber \cite{Bierman:1966zz}. The data are compared to a model of phase space and one-pion exchange. 
The right figure shows data from the LRL bubble-chamber 
experiment where the data are compared to phase space simulations and kinematic reflections \cite{Chinowsky:1968rn}. 
The green boxes mark the range of the $X(2265)$ signal (M$\pm\Gamma/2$).}
\label{Fig:PL_2}
\end{center}
\end{figure*}

%- - - - - - - - - - - - - - - - - - - - - - - - - - - - - - - - - - - - - - - - - - - - - - 
%
Besides the quantitative information discussed in the previous section there are several \pL invariant mass spectra 
that have been published in the last years.
Figures \ref{Fig:PL_1} and \ref{Fig:PL_2} contain a compilation of these spectra at various beam kinetic energies. 
For the production of a state with mass M=2265\,\MeV the threshold beam kinetic energy is $E_{Kin}$=2.18\,GeV. 
Given the large width of the signal it could also be produced at an energy of 
$E_{Kin}$=2.16\,GeV. If the hypothesis is true that a kaonic nuclear cluster is predominantly produced via the $\Lambda(1405)$-doorway scenario \cite{Yamazaki:2002uh,Yamazaki:2007cs}, 
the threshold for the production of a \KNN is $E_{Kin}$=2.35\,GeV (respecting the low mass of the $\Lambda(1405)$ in p+p collisions \cite{Agakishiev:2012xk}). 
\par
The here collected spectra give only a qualitative impression of a potentially visible signal, but a strong deviation of the data from the according models 
in the region of the $X(2265)$-signal (green box) is nevertheless not evident. 
It seems as if there is no hint for a strong visible signal of $X(2265)$ in any of the available data sets which means that if a signal is present its cross 
section can only be a small fraction of the total \pkl final state.  
\par
\begin{figure*}[h]
\begin{center}
\includegraphics[width=0.5\textwidth]{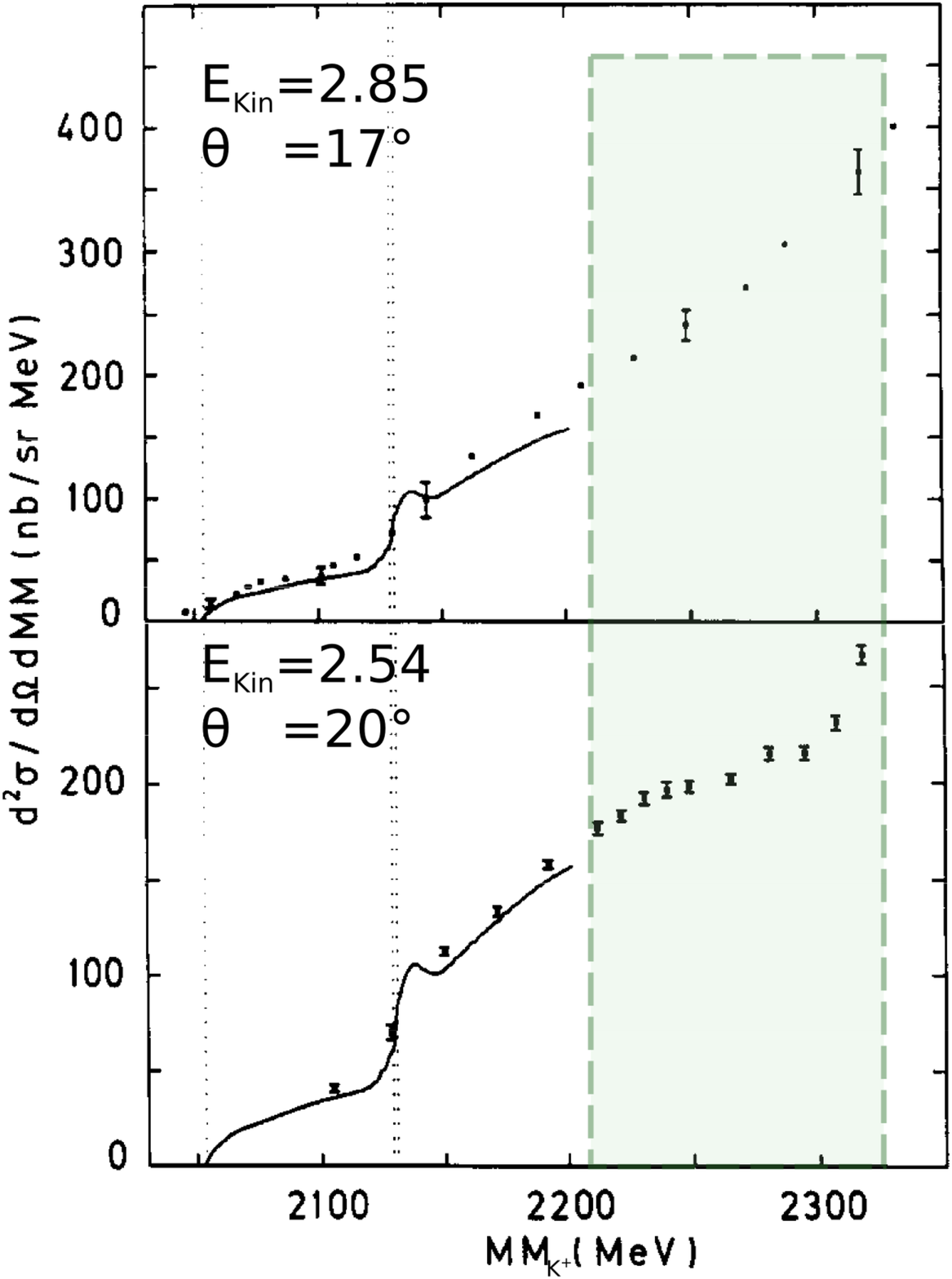}
\caption[]{Two data sets of the missing mass to the $K^{+}$ at beam kinetic energies of 2.85\,GeV, $\theta_{K^{+}}$ = 17$^{\circ}$ (upper) and 2.54\,GeV, $\theta_{K^{+}}$ = 20$^{\circ}$ (lower) \cite{Siebert:1994jy}. The green box marks the range of the $X(2265)$ signal (M$\pm\Gamma/2$).}
\label{Fig:PL_3}
\end{center}
\end{figure*}

There are also other measurements at the same energy as the two DISTO data sets.  
Figure \ref{Fig:PL_3} shows the missing mass to the K$^{+}$ \cite{Reed:1968zza} (at 2.85\,GeV, $\theta_{K^{+}}$ = 17$^{\circ}$) and 
\cite{Hogan:1968zz} (2.54\,GeV, $\theta_{K^{+}}$ = 20$^{\circ}$). 
The data exhibit no significant structure in the indicated $X(2265)$ signal range. One has to note, however, that these are inclusive spectra of 
the production of a residue $X_{R}$ together with a $K^{+}$. 
$X_{R}$ can, depending on the available energy, be composed out of $\Lambda p$, $\Lambda N \pi$, $\Sigma N$ and $\Sigma N \pi$.
While in an exclusive analysis, as done in the DISTO and HADES cases, $X_{R}$ is identified with the \pL system, in Fig. \ref{Fig:PL_3} all 
possible decay channels of the kaonic nuclear cluster 
($YN$ and $YN\pi$) are summarized. 
This is on the one hand a disadvantage, as the background description for the sum of several channels is more difficult 
than in an exclusive analysis, on the other hand, however, this inclusive analysis would compensate for a small branching ratio
of a \KNN in the  \pL channel and a signal would thus nevertheless appear in Fig. \ref{Fig:PL_3}, if kaonic bound states were 
produced with a large cross section. 
Under the assumption of a smooth background underneath the signal this is obviously not the case for both data sets. 
So that also from this point of view it seems that a visible signal by a \KNN cluster is not present in the data.
\par
There is one last data set at 2.83\,GeV, taken by the ANKE collaboration, which contains exclusive \pkl events. Although a proposal had been set up for the analysis of the 
data with respect to the \KNN cluster \cite{ANKEint}, it has not been pursued so far. These are probably the only data whose analysis can quickly resolve the question whether $X(2265)$ 
is due to a physical origin or not. 
%
%- - - - - - - - - - - - - - - - - - - - - - - - - - - - - - - - - - - - - - - - - - - - - - 
\section{Conclusions}
%- - - - - - - - - - - - - - - - - - - - - - - - - - - - - - - - - - - - - - - - - - - - - - 
We have summarized all available experimental measurements of p+p collisions relevant for the search of the 
lightest kaonic nuclear bound state \ppk to cross check the hypothesis that the signal $X(2265)$, reported by the DISTO collaboration, can be associated to a kaonic nuclear bound state \KNNo .
The signal is missing at low ($E_{Kin}<$2.85\,GeV) and high ($E_{Kin}>$2.85\,GeV) beam kinetic energies. 
Its absence can not be explained by a depletion of the $\Lambda(1405)$ yield as was explained with help of Figure \ref{Fig:La1405_CS}. 
The upper limits for the production cross section of $X(2265)$ set at 2.5 and 3.5\,GeV suggest that its contribution to the total \pkl production cross section is only a few percent. 
\par
The strongest argument against the argued nature of $X(2265)$ comes from the method with which the signal was extracted. The deviation 
spectrum technique to search for a new signal is not applicable, if the employed model is not under firm control and if the applied cuts
arbitrarily influence the outcome of the spectra. 
\par
We do, thus, think that the structure $X(2265)$ is very unlikely to be due to a kaonic nuclear bound state. 
On the other hand, the extracted upper limits are still rather sizable and in the order of the predicted yield in Ref.~\cite{Suzuki:2009zze}. 
This calls for new and high statistic experiments to measure the \pkl final state, possibly also employing polarization, and their subsequent analysis with modern
techniques such as a partial wave analysis \cite{Agakishiev:2014dha}. 
The best data set so far for such an analysis is indeed the one measured by DISTO with 400.000 \pkl events (2.85\,GeV). 
Given the fact that N* resonances do play a dominant role in the \pkl final state and given the fact that one data set is not enough to pursue a partial wave analysis 
with a unique solution as shown in Ref.~\cite{Agakishiev:2014dha}, we call for a simultaneous analysis of all available \pkl data sets to finally pin down the issue 
of \KNN production in p+p collisions.  
%
%- - - - - - - - - - - - - - - - - - - - - - - - - - - - - - - - - - - - - - - - - - - - - - - - - - - - - - - - - - - -
\section*{Acknowledgments}
We kindly thank P. Piroue for providing us a nice version of Figure \ref{Fig:DevSpectra_Hogan}. This work has been supported 
by the following grants: DFG EClust 153, DFG FA 898/2-1, BMBF 05P12WOGHH and F\&E-Vertrag TMLFRG1316. 
\bibliography{DISTO_crit}

\end{document}